\documentstyle[preprint,prl,aps]{revtex}
\preprint{IMSc-98/05/23}

\begin{document}
\title{$C\!P$~ Violation and Lifetime Differences of Neutral $B$
Mesons from Correlated $B^0\overline{B^0}$ Pairs}
\author{Nita Sinha 
\ifpreprintsty
\footnote{e-mail: nita@imsc.ernet.in}
\else
\cite{author}
\fi
and Rahul Sinha
\ifpreprintsty
\footnote{e-mail:sinha@imsc.ernet.in}
\else
\cite{author}
\fi
}
\address{ Institute of Mathematical Sciences, Taramani, Chennai 600113, 
India.}
\date{19 May 1998}
\maketitle
\begin{abstract} 
We present a technique to determine the $C\!P$ violating phases, as
well as, the lifetime differences of the mass eigenstates for both
$B_d$ and $B_s$, by considering correlated $B\bar{B}$ pairs produced
at the $\Upsilon$ resonances. We do not require a detailed time
dependent study, but only partial time integrated rates, with the tag
time, either preceding or following the decay of the other $B$ meson
to a final state $f$. $f$ may be a $C\!P$ eigenstate or a non-$C\!P$
eigenstate. 
\end{abstract}

The turn of this century will mark the beginning of a new era in the
study of $C\!P$ violation, the origin of which is not yet
understood. In the standard model, $C\!P$ violation is parametrized by
the Cabibbo-Kobayashi-Maskawa (CKM) matrix\cite{CKM}, and the $B$
meson system is expected to provide a unique testing ground for this
hypothesis.  A large number of dedicated $B$ physics experiments will
start collecting data in the near future and it is expected that
several $B$ parameters will be measured. One of the first to be
operational are the asymmetric $B$ factories at PEP-II and
KEK-B. After two decades of detailed theoretical studies of possible
signals of $C\!P$ violation in $B$ mesons, it is hoped that $C\!P$
violation will finally be seen using time dependent
asymmetries\cite{Asymm} in neutral $B$ mesons. However, few strategies
exist\cite{Width1,Width2,Width3} for the challenging measurement of
the small lifetime difference of the two mass eigenstates.

In this letter we propose a technique to search for $C\!P$ violation,
as well as, measure the lifetime differences of the two neutral $B$
($B_d$ or $B_s$) mass eigenstates. We consider coherent $B^0
\overline{B^0}$ pairs produced at $\Upsilon$ resonances, where, one of
the neutral $B$ meson decay acts as a flavour tag (which for the
purpose of illustration we take to be a semileptonic decay), while the
other, decays to a hadronic state $f$ or $\bar{f}$ ({\it i.e.},
$\Upsilon\to B^0\overline{B^0}\to l^\pm f(\bar{f}) X^\mp$). The
hadronic state $f$ is chosen to be one, into which both $B^0$ and
$\overline{B^0}$ can decay.  In the absence of $C\!P$ violation, the
number of such events with an $l^+$ and $f$ or $\bar{f}$ in the final
state, equals those with an $l^-$.  In addition for equal lifetimes,
if we add the number of events to $f$ and ${\bar f}$, tagged by the
same charge lepton, then the number of events where the semileptonic
decay precedes the hadronic, equals those in which the semileptonic
decay follows the hadronic.  Using these facts, we construct certain
ratios of number of events, which in the absence of $C\!P$ violation
and/or equal lifetimes, should each be identically half.  Such ratios
can be obtained by comparing the number of events where the
semileptonic decay is selectively time ordered with respect to the
hadronic decay, to that where all such events are added irrespective
of the time ordering.  Hence, the deviation of these ratios from half
would provide a remarkable signal of $C\!P$ violation and/or unequal
lifetimes. Some of these ratios are unique, as these cannot be recast
into asymmetries of particle--antiparticle events, as is done
conventionally.

In the case of $C\!P$ eigenstates $f$ ({\it e.g.} $B_d\to J\!/\!\psi
K_s$), these ratios enable us to determine not only the $C\!P$
violating weak phase but most importantly, the width difference
($\Delta\Gamma$) of the neutral $B$ mass eigenstates as
well. Non-$C\!P$ eigenstates where $f$ and $\bar{f}$ are identifiable
({\it e.g.} $B_d\to D^{\pm}\pi^\mp$), also allow the determination of
the width difference along with the weak phase, strong phases and the
amplitudes involved. In cases where the final states $f$ and $\bar{f}$
cannot be separated ({\it e.g.} $B_d\to D^0(\overline{D^0})\pi^0$), our
approach can be particularly useful since we add $f$ and $\bar{f}$
events. However, in such a case the width difference cannot be
determined, in general, from the same mode.

Calculation of the box--diagrams for $B^0-\overline{B^0}$
mixing\cite{Box} leads to a tiny estimate for the width difference in
$B_d$, of the order $\Delta\Gamma<0.01\;\Gamma$. However, for $B_s$
meson, the width difference can be larger, $\Delta\Gamma\sim
0.1\;\Gamma$\cite{Width1,Width2}. Recent data\cite{CLEO} on exclusive
two-body $B$ decays, indicates significant gluonic penguin
amplitudes. Sizable contribution to $\Delta\Gamma$, via the absorptive
part of dipenguin diagrams\cite{dipenguin} can thus be expected. This
contribution has not been included in the above estimates for
$\Delta\Gamma$. Hence, one should not preclude a measurable width
difference even in the case of $B_d$ mesons.

The decay amplitude for a coherent $B^0 \overline{B^0}$ state , with a $B$
decaying to $l^+X^-$ at time $t_0$, and the other B decaying to $f$ at
time t is given by \cite{BigiSanda},
\begin{eqnarray} 
\displaystyle A[(l^+)_{t_0},(f)_{t}]&=&
\displaystyle <l^+X^-|H|B^0(t_0)> <f|H|\overline{B^0}(t)>\nonumber \\ &&
+\displaystyle (-1)^C <l^+X^-|H|\overline{B^0}(t_0)> <f|H|{B^0}(t)>,
\label{upsil}
\end{eqnarray}
where, $C$ is the charge parity of the  $B^0 \overline{B^0}$ state. 
The hadronic matrix elements for pure $B^0$ and $\overline{B^0}$ may be
written as,
\begin{eqnarray}
<f|H|B^0> &=& A_1 e^{i\phi_1} e^{i\delta_1}\nonumber \\
<f|H|\overline{B^0}> &=&  A_2 e^{i\phi_2} e^{i\delta_2} \nonumber \\
<\bar{f}H|B^0> &=&  A_2 e^{-i\phi_2} e^{i\delta_2}\nonumber \\
<\bar{f}H|\overline{B^0}> &=& A_1 e^{-i\phi_1} e^{i\delta_1},
\end{eqnarray}
where, $\phi_1$, $\phi_2$ are the weak phases, $\delta_1$,
$\delta_2$ are the strong phases, and  $A_1$, $A_2$ are the
magnitudes of the amplitudes respectively. We restrict our analysis to
final states with just one weak phase in each of the above.
Using, the time evolution equations for initially pure $B^0$ and
$\overline{B^0}$ states, we may rewrite Eq.(\ref{upsil}),  as well as
those corresponding to, $l^-f,l^+\bar{f},
l^-\bar{f}$ in the
final state as,
\begin{eqnarray}
A[l^+f ]&=&\displaystyle\xi F  A_1 e^{i\phi_1} e^{i\delta_1} ( C_+ +C_-
 r|\xi|^{-1}
  e^{i\phi} e^{i\delta}) \nonumber\\
 A[l^-f ]&=&-\displaystyle \bar{F}  A_1 e^{i\phi_1} e^{i\delta_1} ( C_- +C_+
 r|\xi|^{-1} 
  e^{i\phi} e^{i\delta}) \nonumber\\
 A[l^+\bar{f} ]&=&\displaystyle  F  A_1 e^{i\phi_1} e^{i\delta_1} (
 C_+r |\xi| e^{-i\phi} e^{i\delta} + C_-) \nonumber \\
 A[l^- \bar{f} ]&=&-\displaystyle \bar{F}\xi^{-1}  A_1 e^{i\phi_1}
 e^{i\delta_1} ( C_-r |\xi|e^{-i\phi} e^{i\delta} + C_+), 
\label{amps}
\end{eqnarray}
where,
\begin{eqnarray}
&&r=A_2/A_1,\;\phi=\phi_2-\phi_1-2 \beta,\; 
\delta=\delta_2-\delta_1, \nonumber\\
&&C_\pm=g_+(t_0)g_\mp(t)+(-1)^C 
g_-(t_0)g_\pm(t),\;\; F(\bar{F})=<l^\pm X^\mp|H|B^0(\overline{B^0})>,
\nonumber \\ 
&&g_{\pm}=\displaystyle e^{-(\frac{\Gamma}{2}+i
M)t}\left\{\cos\!\left[(\Delta  
M-i \frac{\Delta \Gamma}{2})\frac{t}{2}\right],i\sin\!\left[(\Delta
M-i \frac{\Delta  
\Gamma}{2})\frac{t}{2}\right]\right\},\nonumber
\end{eqnarray}
with $|F|^2=|\bar{F}|^2$; $\xi=\displaystyle|\frac{p}{q}|e^{2 i
\beta}$, where the parameters $p$ and $q$ relate\cite{Quinn} the
flavor eigenstates to the mass eigenstates. 

Note that all the amplitudes in the above equation, involve the same
weak phase $\phi$. All interference terms involving this phase are
washed out for the $\Upsilon$ decays (C=1), if we evaluate the time
integrated rates.  In what follows we restrict our discussion to
$C=1$ case only. While detailed time dependent studies can always be
used to determine the $C\!P$ violating phase and the finite width
difference, during the initial stages of operation, limited statistics
may not allow one to do so. However, it will still be possible to
obtain information on these, by partial time integrated rates. We
implement this by choosing the decay into one of the $l^\pm$ or
$f$($\bar{f}$) states to take place at any time, while the decay to
the other state occurs either before or after it, $ {\it i.e.}$ the
decay times are ordered with respect to each other.

We consider the sum of events where the semileptonic decays to  $l^+$ 
precedes the hadronic decay and that to $l^-$ occurs after the hadronic 
decay. We call such events opposite time events. Similarly, the like 
time events are defined to be those where both $l^+$ and $l^-$ decays 
precede the hadronic decays. A remarkable signal of $C\!P$ violation  or 
finite lifetime difference is obtained by taking the ratio of the 
opposite time or like time events respectively, to the total number of 
events. The advantage of considering ratios of number of events, is
that systematic errors tend to cancel.  

Using Eq.(\ref{amps}) the two  ratios can be written as,
\begin{eqnarray}
R_1&=&\frac{N_<(l^+f)+N_<(l^+\bar{f})+N_>(l^-f)+N_>(l^-\bar{f})}{N
(l^+f)+N(l^+\bar{f})+N(l^-f)+N(l^-\bar{f})}=\frac{1}{2}(1+\epsilon_1)
\nonumber\\
&=&\frac{1}{2}[1-\Bigl(
(1-y^2)\frac{x}{1+x^2}\sin\phi+
\frac{a_{ll}}{2}\;y\cos\phi\Bigr)\frac{2\;r}{1+r^2}\cos\delta]
\label{ratio1}\\
R_2&=&\frac{N_<(l^+f)+N_<(l^+\bar{f})+N_<(l^-f)+N_<(l^-\bar{f})}{N
(l^+f)+N(l^+\bar{f})+N(l^-f)+N(l^-\bar{f})}=\frac{1}{2}(1+\epsilon_2)
\nonumber\\
&=&\frac{1}{2}[1-\Bigl(\frac{a_{ll}}{2}\;(1-y^2)\frac{x}{1+x^2}\sin\phi+
y\cos\phi\Bigr)\frac{2\;r}{1+r^2}\cos\delta]
\label{ratio2}    
\end{eqnarray}
where, $N_<$ denotes the number of events where the hadronic decay
precedes the semileptonic, {\it i.e.} $t < t_0$, while $N_>$ denotes
the number of events where the hadronic decay occurs after the
semileptonic, {\it i.e.} $t > t_0$; $N$ without the subscript
represents the integrated numbers, where each of $t_0$ and $t$ runs
from $0$ to $\infty$. Also $x=\displaystyle\frac{\Delta M}{\Gamma}$,
$y=\displaystyle\frac{\Delta \Gamma}{2\Gamma}$, $\Delta M$ and $\Delta
\Gamma$ denote the mass and width differences of the two $B$ mass
eigenstates respectively, while $\Gamma$ is the average width of the
two states. In the above we have used the following integrals,
\begin{eqnarray}
\int_0^{\infty} dt_0 \int_0^{t_0} dt |C_+|^2 &=&\int_0^{\infty} dt_0
\int_{t_0}^{\infty} dt |C_+|^2=\frac{x^2+y^2}{4 \Gamma^2 
(1+x^2) (1-y^2)}\nonumber\\
\int_0^{\infty} dt_0 \int_0^{t_0} dt |C_-|^2&=& \int_0^{\infty} dt_0
\int_{t_0}^{\infty} dt |C_-|^2=\frac{2+x^2-y^2}{4 \Gamma^2 
(1+x^2) (1-y^2)}\nonumber\\
\int_0^{\infty} dt_0 \int_0^{t_0} dt
C_{\mp}C_{\pm}^*&=&-\int_0^{\infty} dt_0 \int_{t_0}^{\infty} dt
C_{\mp}C_{\pm}^*=\frac{-y(x^2+1)\pm i x(1-y^2)}{4 \Gamma^2
(1+x^2) (1-y^2)},
\end{eqnarray}
and we have related $|\xi|$ to the charge asymmetry in like sign
dilepton events defined by,
\begin{equation}
a_{ll}=\frac{N(l^+l^+)-N(l^-l^-)}{N(l^+l^+)+N(l^-l^-)}=
\frac{|\xi|^4-1}{|\xi|^4+1}, 
\end{equation}
implying,
\begin{equation}
|\xi|\pm\frac{1}{|\xi|}=\frac{(1+a_{ll})^{1/2}\pm(1-a_{ll})^{1/2}}
{(1-a_{ll}^2)^{1/4}}\approx \{2,a_{ll}\}(1+O(a_{ll}^2)+...).
\label{xi}
\end{equation}
As is well known, $a_{ll}$ is a signature of direct $C\!P$ violation.
In the standard model $a_{ll}$ is expected to be very small $\approx
10^{-3}-10^{-4}$. Upper bound for $B_d$ is
$|a_{ll}|<0.18$\cite{cleo,pdg}. While our discussions can be
generalized for arbitrary $a_{ll}$, we take $a_{ll}$ to be small,
retaining, in Eq.(\ref{xi}), only linear order terms in $a_{ll}$.
Expressing $|\xi|$ in terms of the observable $a_{ll}$, has the
advantage that one makes no assumption regarding the size of ${\rm
  Im}\epsilon_B$ or ${\rm Re}\epsilon_B$\cite{xing} . An exchange of
$l^+ \leftrightarrow l^-$ or $N_< \leftrightarrow N_>$ in the ratios
$R_1$ and $R_2$ results in switching the sign of $\epsilon_1$ and
$\epsilon_2$. It may be pointed out that any deviation of $R_1$ from
half is an unambiguous signature of $C\!P$ violation. $R_1$ and $R_2$
are constructed such that $R_1$ measures $\sin\phi$, whereas $R_2$
measures $y$, both with correction terms of the order $a_{ll}$.

We can also obtain the following unusual but remarkable signals of:
{\it i} )
$C\!P$ violation in the limit $y\to 0$, {\it ii} ) non-vanishing width
difference in the limit $\phi\to 0$,
\begin{eqnarray}
R_{3}&=&\frac{N_<(l^{+}f)+N_<(l^{+}\bar{f})}
{N(l^{+}f)+N(l^{+}\bar{f})}=\frac{1}{2}(1+\epsilon_3)\nonumber\\
&=&\displaystyle\frac{1}{2}[1- 
\frac{2 r}{1+r^2}\cos\delta\Bigl(\frac{x}{1+x^2}(1-y^2)\sin\phi+
y\cos\phi\Bigr) 
(1+\displaystyle\frac{1}{2}a_{ll})(1+\displaystyle\frac{
x^2+y^2}{2(1+x^2)}a_{ll})^{-1}] \\
R_{4}&=&\frac{N_<(l^{-}f)+N_<(l^{-}\bar{f})}
{N(l^{-}f)+N(l^{-}\bar{f})}=\frac{1}{2}(1+\epsilon_4)\nonumber\\
&=&\displaystyle\frac{1}{2}[1+ 
\frac{2r}{1+r^2}\cos\delta\Bigl(\frac{x}{1+x^2}(1-y^2)\sin\phi-
y\cos\phi\Bigr) 
(1-\displaystyle\frac{1}{2}a_{ll})(1-\displaystyle\frac{
x^2+y^2}{2(1+x^2)}a_{ll})^{-1}]. 
\label{ratio3}
\end{eqnarray}
The above ratios are unique, as the corresponding deviations from half
({\it i.e. }$\epsilon_3,\epsilon_4\neq 0$), cannot be rewritten in
terms of an asymmetry between particle and antiparticle events, but
only as an asymmetry of $t<t_0$ and $t>t_0$ events. Only if
$\Delta\Gamma=0$, as well as $|\xi|=1$, $R_3$ and $R_4$ can be
rewritten in terms of asymmetries of the usual type. It may be pointed
out that $R_3$, $R_4$ are linearly independent of $R_1$, $R_2$ only
for $|\xi|\neq 1$. For $a_{ll}=0$ and $y=0$, we get,
$\epsilon_1=\epsilon_3=-\epsilon_4$ and $\epsilon_2=0$. We wish to
point out that $R_1,\cdots,R_4$ do not exhaust the list of independent
ratios.  When $f$ and $\bar{f}$ are distinguishable, then the ratios
formed by subtracting $f$ and $\bar{f}$ events, corresponding to each
of $R_1,\cdots,R_4$ can also be written down.

\underline{\it $C\!P$ eigenstate $f$}: If $f$ is a $C\!P$ eigenstate,
$\delta=0$, $r=n_f$ (where $n_f=\pm 1$, is the $C\!P$ parity of
$f$). Hence, in principle, if $x$ and $a_{ll}$ are known, the $C\!P$
violating weak phase $\phi$ and the width difference related to $y$,
can be determined from the ratios $R_1$ and $R_2$; $\sin\phi$ and
$y^2$ are the solutions of the following cubic equations:
\begin{eqnarray}
4\sin^3\phi+n_f\frac{2 (2\epsilon_1-a_{ll}\epsilon_2)}{
x_0}(\sin^2\phi-1)+(4\epsilon_2^2+\epsilon_1^2 a_{ll}^2-4
a_{ll}\epsilon_1\epsilon_2-4)\sin\phi=0\label{sinphi}\\ 
(1-y^2)^2(4 y^2-\epsilon_1^2 a_{ll}^2-4\epsilon_2^2+4\epsilon_1 \epsilon_2
a_{ll})-\frac{1}{x_0^2}(4\epsilon_1^2+\epsilon_2^2 a_{ll}^2-4
a_{ll}\epsilon_1\epsilon_2)y^2=0\label{y}  
\end{eqnarray}
where, $x_0=\displaystyle\frac{x}{1+x^2}$.

Experimentally $x$ is determined\cite{pdg} using, either $\Delta
M$, measured by fits to the time dependent oscillation probabilities of
a created $B^0$ decaying as a $B^0({\bar B^0})$, or from $\chi$, the
integrated probability of the same. Both these  determinations already
assume $\Delta\Gamma=0$. In the ratios $R_{1-4}$, the value of
$x$ could be rewritten in terms of $y$ and the observable
$\chi$, given by, $\chi=\displaystyle\frac{x^2+y^2}{2(1+x^2)}$.
Equations analogous to Eq.(\ref{sinphi},\ref{y}), for $\sin\phi$ and
$y$ can also be written in terms of $\chi$, instead of $x_0$.

For $B_d$, it is expected that $y^2\ll 1$; the above relations for
$\sin\phi$ and $y$, then simplify to:
\begin{eqnarray}
&&\sin\phi=n_f\frac{a_{ll}\epsilon_2-2\epsilon_1}{2 x_0}\\
&&y=n_f\frac{(a_{ll}\epsilon_1-2\epsilon_2)x_0}{
\sqrt{4 x_0^2-4\epsilon_1^2-\epsilon_2^2 a_{ll}^2+4 a_{ll}\epsilon_1
\epsilon_2}}. 
\end{eqnarray}
In the limit $a_{ll}\to 0$, a signal of finite width difference is a
nonzero value of $\epsilon_2$. We wish to emphasize that in this
limit, {\em the determination of $\Delta\Gamma$ is independent of
$x$}. Using $B_d\to J\!/\!\psi K_s$, in a sample of $10^7\Upsilon $'s
one can constrain $\sin (2\beta)<0.32$ at $3 \sigma$ (for
$a_{ll}=0,y^2\ll 1$). Hence, it is possible to measure $\beta$ over
the entire currently allowed range. However, the same sample will
provide a very weak constraint on the width difference. Combining
$B_d\to J\!/\!\psi K_s$, $J\!/\!\psi K^*$ and $\psi(2 S)K^*$ (with
$K^*$ seen in the $K_s \pi^0$ mode), to increase statistics, in a
sample of $10^8 \Upsilon $'s, we obtain a sensitivity
$\Delta\Gamma_d=(0.06,0.16)\Gamma_d$ for $\beta =(10^o,35^o)$ at
$3\sigma$ (for $a_{ll}=0$). Here we have assumed that one $C\!P$
parity state is dominant in the $J\!/\!\psi K^*$ mode\cite{CLEO2}. If
separate $C\!P$ parity states exist, they can still all be added,
weighted by the corresponding $C\!P$ parities\cite{Kayser}.  Once
$\phi$ is measured for various modes, the events for all the modes can
be combined to determine $y$ by introducing a weight factor of
$1/\cos\phi$ corresponding to each mode.

In the $B_s$ case, one anticipates $x_0\ll 1$, moreover modes governed
by $b\to c \bar{c}s$ (e.g. $B_s\to J\!/\!\psi\phi$) are expected to
have a tiny weak phase, $\it{i.e.}$, $\phi\approx 0$. These modes can
hence be used to determine the width difference for $B_s$ through the
relation, \begin{equation} y^2=\epsilon_2^2-\epsilon_1^2.
\end{equation} The above relation is {\em independent of the value of
$|\xi|$ as well as $x$}. In fact, $|\xi|$ itself can be expressed as,
\begin{equation}
|\xi|=\sqrt{\frac{\epsilon_1+\epsilon_2}{\epsilon_2-\epsilon_1}}.
\end{equation} 
A comparison of the value of $|\xi|$ determined through
this relation and that evaluated using the charge asymmetry in like
sign dilepton events, would provide a useful consistency check. In the
limit $a_{ll}\to 0$ and $\phi\approx 0$, one gets,
$y=n_f\epsilon_2$. In such a case the sign of $y$ can also be
determined. To obtain sensitivity of $\Delta\Gamma_s=0.2\Gamma_s$
would require 900 $B_s\to J\!/\!\psi\phi$ tagged events produced at
$\Upsilon(5S)$. The statistics can be significantly improved by
combining the various modes involving the transition $b(\bar{b})\to c
\bar{c}s(\bar{s})$.

\underline{\it Non--$C\!P$ eigenstate $f$}: When it is experimentally
feasible to distinguish $f$ and $\bar{f}, ${\it e.g.} $B_d\to D^\pm
\pi^\mp$, there is in principle enough information to extract all the
unknown variables, including the weak phase and the width
difference. This can be achieved by measuring the ratios
$R_1,\cdots,R_4$ and the analogous ratios where the $f$ and $\bar{f}$
events are subtracted, giving a total of eight observables. The
asymmetry, for a typical single mode {\it e.g.} $D^- \rho^+$, is $\sim
0.005\sin(2\beta+\gamma)$ (using $a_{ll}=0, y^2\ll 1$ and
factorization approximation). However, all modes involving the
transition $b(\bar{b})\to c \bar{u} d (\bar{c}u \bar{d})$, can be
combined to improve statistics; the total branching ratio of such
modes is of the order of a few percent, allowing for a possible
observable asymmetry\cite{Dunietz}.  The decay mode $B_d\to D K_s$ is
calculated to have a large asymmetry $\sim 0.3 \sin(2\beta+\gamma)$ in
the factorization approximation.  The constraint,
$(2\beta+\gamma)<55^o$ at $3\sigma$ can be obtained from 146 $B_d \to
D K_s$ tagged events produced at $\Upsilon(4S)$ (assuming
$a_{ll}=0,y^2\ll 1$). One of the difficulties encountered in $C\!P$
violation studies involving a neutral $D$ in the final state, {\it
e.g.} $B_d\to D^0 \pi^0, B_s\to D^0 K_s$, is that it is hard to
identify the flavour of the $D$ meson. In our technique, this poses no
problem for a finite width difference, as the final states $f$ and
$\bar{f}$ can be added. If $a_{ll}\neq 0$ and $y\neq 0$, the four
independent ratios $R_1,\cdots,R_4$ can be used to determine, the weak
phase, the width difference and the unknown product $r/(1+r^2)
\cos\delta$. However, in case $a_{ll}=0$ but $y\neq 0$, $\phi$ can be
determined only with prior knowledge of $y$.
  
To conclude, we present a simple technique to measure both the $C\!P$
violating weak phases as well as width differences of the two neutral
$B$ mass eigenstates using coherent $B_d \overline{B_d}$ or $B_s
\overline{B_s}$ pairs, produced at the $\Upsilon$ resonances. This is
achieved, without a detailed time study, considering only partial time
integrated rates where the tag time is selectively ordered with
respect to the decay time of the other $B$ decaying hadronically to
$f$.  We construct certain ratios which signal $C\!P$ violation and/or
a lifetime difference, if different from half.  Observation of such
ratios can be used to cleanly measure the weak phase $\phi$ and the
width difference $\Delta\Gamma$ simultaneously. The determination of
the $\Delta\Gamma$ is independent of a measured value of the mass
difference $\Delta M$.  This technique is applied to the case where
$f$ is a $C\!P$ eigenstate such as in $J\!/\!\psi K_S$, as well as for
a non-$C\!P$ eigenstate, {\it e.g.} $D^0 K_s$. If a measurable width
difference exits, our approach offers a particular advantage for final
states involving a neutral $D$, since we can add $D^0$ and
$\overline{D^0}$ events, which may not always be distinguishable.

We are very grateful to Prof. L. M. Sehgal for valuable suggestions.
We also thank Prof. Tariq Aziz and Prof. G. Rajasekaran for
discussions.

\end{document}